\documentclass[draft]{agujournal2019}
\usepackage{url} 
\usepackage{lineno}
\usepackage[inline]{trackchanges} 
\usepackage{soul}
\draftfalse

\journalname{ Journal of Advances in Modeling Earth Systems (JAMES)}

\begin{document}

%
%


\title{Indicator patterns of forced change learned by an artificial neural network}

%
%




\authors{Elizabeth A. Barnes\affil{1}, Benjamin Toms\affil{1}, James W. Hurrell\affil{1}, Imme Ebert-Uphoff\affil{2,3}, Chuck Anderson\affil{4,5}, and David Anderson\affil{5}}


\affiliation{1}{Department of Atmospheric Science, Colorado State University, Fort Collins, CO}
\affiliation{2}{Cooperative Institute for Research in the Atmosphere, Colorado State University, Fort Collins, CO}
\affiliation{3}{Department of Electrical and Computer Engineering, Colorado State University, Fort Collins, CO}
\affiliation{4}{Department of Computer Science, Colorado State University, Fort Collins, CO}
\affiliation{5}{Pattern Exploration LLC, Fort Collins, CO}





\correspondingauthor{Elizabeth A. Barnes}{eabarnes@rams.colostate.edu}




\begin{keypoints}
\item Indicator patterns of forced change of temperature and precipitation are identified in climate models using artificial neural networks
\item These patterns are episodic in time and are present within observations, offering insights into observed indicators of forced change
\item Neural network visualization tools can be leveraged to assess regional differences between simulated and observed indicators of change
\end{keypoints}

%
%

%
%


\begin{abstract}
Many problems in climate science require the identification of signals obscured by both the  ``noise'' of internal climate variability and differences across models. Following previous work, we train an artificial neural network (ANN) to identify the year of input maps of temperature and precipitation from forced climate model  simulations. This prediction task requires the ANN to learn forced patterns of change amidst a background of climate noise and model differences. We then apply a neural network visualization technique (layerwise relevance propagation) to visualize the spatial patterns that lead the ANN to successfully predict the year. These spatial patterns thus serve as ``reliable indicators'' of the forced change. The architecture of the ANN is chosen such that these indicators vary in time, thus capturing the evolving nature of regional signals of change. Results are compared to those of more standard approaches like signal-to-noise ratios and multi-linear regression in order to gain intuition about the reliable indicators identified by the ANN. We then apply an additional visualization tool (backward optimization) to highlight where disagreements in simulated and observed patterns of change are most important for the prediction of the year. This work demonstrates that ANNs and their visualization tools make a powerful pair for extracting climate patterns of forced change.
\end{abstract}

\section{Plain Language Summary}
Many problems in climate science require the identification of signals obscured by both the ``noise'' of internal climate variability and differences across models. Here, we demonstrate that machine learning methods, specifically artificial neural networks and their visualization tools, can be used to visualize indicators of change in surface temperature and precipitation within climate models and the observations. This work demonstrates that ANNs and their visualization tools make a powerful pair for extracting climate patterns of forced change.

%
%

%


%
%
%
%

\section{Introduction}
Climate science has often required the identification of signals obscured by both climate ``noise'' and disagreements across models, and the field has a rich history of tools developed for this purpose. In addition to a large number of standard statistical techniques (Zwiers \& von Storch, 2004), a common recent approach has been the utilization of large ensembles of climate model simulations (Deser et al. 2012; Hawkins et al., 2016; Kumar \& Ganguly, 2018; Lehner et al., 2016). In particular, this approach allows researchers to estimate the climate ``noise'', defined as the range of climate outcomes arising from unpredictable internal (or natural) climate variability under a particular radiative forcing scenario, and the structural component of uncertainty due to model differences when multi-model ensembles are used (Deser et al. 2020). Moreover, the forced climate signal associated with a radiative forcing  scenario can be obtained by averaging across a sufficient number of ensemble members, since time sequences of internal variability are randomly phased between individual ensemble members. While the resulting ensemble-mean spatial pattern captures the forced response,  it is difficult to identify this pattern in a single year of observations because the climate of any given year is always a combination of the forced signal and internal variability.

The challenge of identifying the forced response in a single realization of the climate system has been recently approached with a variety of advanced statistical techniques. For example, Sippel et al. (2019) and Wills et al. (2020) employ novel dynamical adjustment techniques and optimal filtering, respectively, to extract the full forced response from that of internal variability within a single ensemble member of a single climate model. Another approach to identify climate signals was recently demonstrated by Barnes et al. 2019 (hereafter B19). They showed that machine learning techniques, specifically artificial neural networks (ANNs), are powerful and useful tools that can help identify patterns of forced climate change within climate model simulations as well as observations. This was achieved by successfully training an ANN to predict the year of a given annual-mean temperature (or precipitation) map from forced CMIP5 simulations. Since each model simulation differs in the internal variability of any given year, this design requires the ANN to learn reliable indicator patterns of each year amidst a background of internal variability and model disagreement. These indicator patterns are thus a combination of the common forcings (e.g. aerosol emissions, anthropogenic greenhouse gas) across all simulations. The climate response to external forcings is typically computed as the average change (or trend) in time across many climate model simulations. In contrast, the indicator patterns identified by the ANN offer the most reliable regions for identifying change in any given year, taking into account the regional internal variability, signal, and disagreement across models. These patterns may thus be used to detect and attribute observed regional change to external forcings, or to identify where climate model biases are most important for obscuring regional change.

While B19 demonstrated that ANNs are capable of identifying forced patterns of change in a single annual-mean map of temperature or precipitation, they did not present the patterns themselves due to the complexity of visualizing the decision-making process of a nonlinear ANN. Instead, they showed oversimplified patterns that came from a much simpler ANN. Here, we apply a recently developed neural network visualization tool (layerwise relevance propagation) to explore the ANN's indicator patterns in detail and quantify how they may vary nonlinearly in time. We compare the patterns from the ANN with those obtained from more classical approaches (e.g. signal-to-noise ratios and multi-linear regression) to gain further intuition about the ANN output. Finally, we apply an additional neural network visualization tool (backward optimization) to map the regions where climate model biases may be most relevant when identifying forced change. 

\section{Data}
\subsection{CMIP5 climate model output}
We analyze the same data used in B19. Namely, annual-mean global two-meter air temperature and precipitation rate output from climate model simulations performed for the Coupled Model Intercomparison Project, phase 5 (CMIP5; Taylor et al., 2012). Due to data availability, single simulations from 29 models are analyzed for temperature, while 22 models are analyzed for precipitation (see Supp. Tables 1 and 2). The ANN requires all input maps to be the same size; thus,  prior to analysis, all fields were interpolated to a common 4 degree latitude by 4 degree longitude grid (45 latitude values by 90 longitude values = 4050 total grid points). The small number of grid points in this relatively coarse grid helped substantially reduce the time required for ANN training. 
 
We analyze annual-mean temperature and precipitation under historical forcing (from 1920 through 2005) and then the RCP8.5 scenario through the year 2099 (Meinshausen et al., 2011). Since all of the model simulations have similar external forcings, deviations across model projections mostly reflect differences due to climate model physics, resolution, and numerics (i.e., model uncertainty) as well as differences in the unforced, or internal, variability of the climate system (Hawkins \& Sutton, 2009; Lehner et al., 2020).  

\subsection{Observations}
We assess the applicability of the ANN trained on climate models to the real world by evaluating the ANN's skill in predicting the year of observed maps of annual mean temperature and precipitation.
For observations of surface temperature, we utilize the BEST (Berkeley Earth Surface Temperature) gridded fields from Berkeley Earth (Rohde et al., 2013). Specifically, we analyze the Monthly Land $+$ Ocean, Average Temperature with Air Temperatures at Sea Ice (name on website given as Recommended; 1850 - Recent) interpolated to a common grid of 4 degree latitude by 4 degree longitude. The climatology field for each month is provided by BEST and was added to the BEST monthly anomalies to obtain the total temperature (K). Data coverage is incomplete in BEST prior to the mid 20th Century. We thus only analyze data from 1956-2018 when there is complete global coverage. 

Monthly observational precipitation fields were obtained from the NOAA Global Precipitation Climatology Project (GPCP), version 2.3, for 1979-2018 (Adler et al., 2018). Data from rain gauge stations, satellites, and sounding observations are merged in GPCP to estimate monthly rainfall (mm/day). Data were downloaded from the NOAA ESRL website (see Acknowledgements) and were interpolated to the common 4 degree grid prior to analysis.

\section{Neural network methods}
\subsection{Neural network architecture}
In B19 the analysis was set up as a prediction problem. Annual-mean maps of temperature (or precipitation) were taken as input and an ANN was trained to predict the year of the map, as shown in Supp. Figure 1. Specifically, each grid point in the input map was represented by a unit in the input layer of the ANN (4050 input neurons in total from the 45 latitude by 90 longitude grid points). The input layer was followed by a number of hidden layers, and the final output layer was a single neuron,  representing  the yearly prediction as a single scalar. This type of set-up is known as a regression task, since the output was a continuous number. 

In contrast, in this work we frame the prediction problem as a classification task; namely, rather than generating an estimate of the year as a continuous number, we instead estimate which one of a number of possible classes the year belongs to. Specifically, the output layer of the ANN in Figure 1 consists of 22 classes, each one representing one decade, and it is the ANN's task to determine which class (i.e. decade) the input map belongs to. Formulating the problem as a classification task is a necessity because the specific ANN visualization tool we employ (layerwise relevance propagation (LRP); Sec. 3.3) was developed for classification architectures, not regression architectures. 

 \begin{figure}
 \begin{center}
 \includegraphics[width=\textwidth]{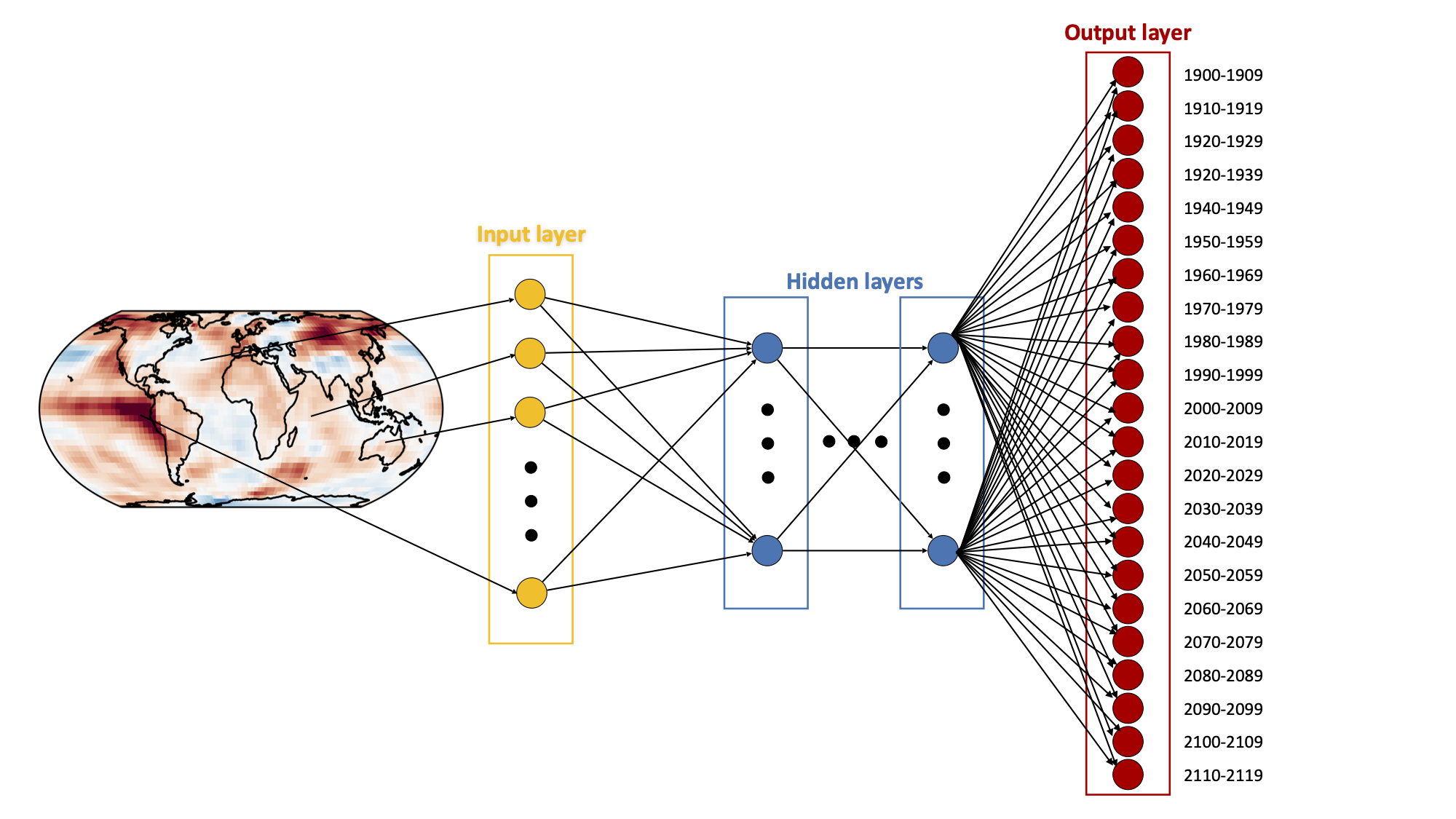}
 \end{center}
 \caption{Schematic of ANN architecture employed here to predict the year of a map of 2-meter temperature. The output layer is divided into classes, each representing a single decade. The ANN task is to predict the class probabilities associated with the input, which is called a classification task.  Here fuzzy classification is used to encode the specific year, and binary cross-entropy is used during training.}
 \end{figure}

ANNs used for classification typically use crisp encoding (i.e. one-hot encoding) for the output classes, mapping the year of an input sample to exactly one output class. For example, the year 1920 would be encoded as completely belonging to the class 1920-1929, and no other class.  This results in large information loss since there is no information left on whether 1920 lies toward the beginning, middle, or end of that decade, or whether neighboring years share similar characteristics. To retain such information we instead use fuzzy encoding, which maps any year to one or more neighboring classes with varying degrees of membership (encoded as probabilities), with the sum of the probabilities summing to one (Zadeh, 1965).  Using triangular membership functions (Zadeh, 1965) with a width equal to one decade results in each year being mapped to one or two neighboring classes with non-zero probabilities. Specifically, if one denotes each output class by its central year, e.g. 1935 for 1930-1939, then the class probabilities are chosen such that the decade-weighted sum equals the exact year.  This encoding and decoding is visualized in Figure 2, where the decade classes are specified on the y-axis, and the corresponding probabilities associated with each class are specified on the x-axis. For each colored year (1925, 1984, 2040, 2078), the dots in the same color indicate the corresponding probabilities. For example, the year 1925 is encoded as a single probability of 1.0 for the class called ``1925'', while the year 1984 has a probability of 0.9 for class ``1985'' and probability 0.1 for class ``1975''. Indeed, the decade-weighted sum, $0.9\cdot1985 + 0.1\cdot1975 = 1984$, gives the correct year of 1984. This approach implements ``fuzzy decadal classification'' at the ANN output layer and the ANN is then tasked with assigning the correct (fuzzy) probabilities for an input sample to each of these classes/decades. This multi-label, fuzzy classification approach allows for encoding of the exact (true) year in the output classes, while still ensuring that the output is a set of class probabilities for use with our preferred visualization tool, LRP (Section 3.3).

 \begin{figure}
 \begin{center}
 \includegraphics[width=\textwidth/2]{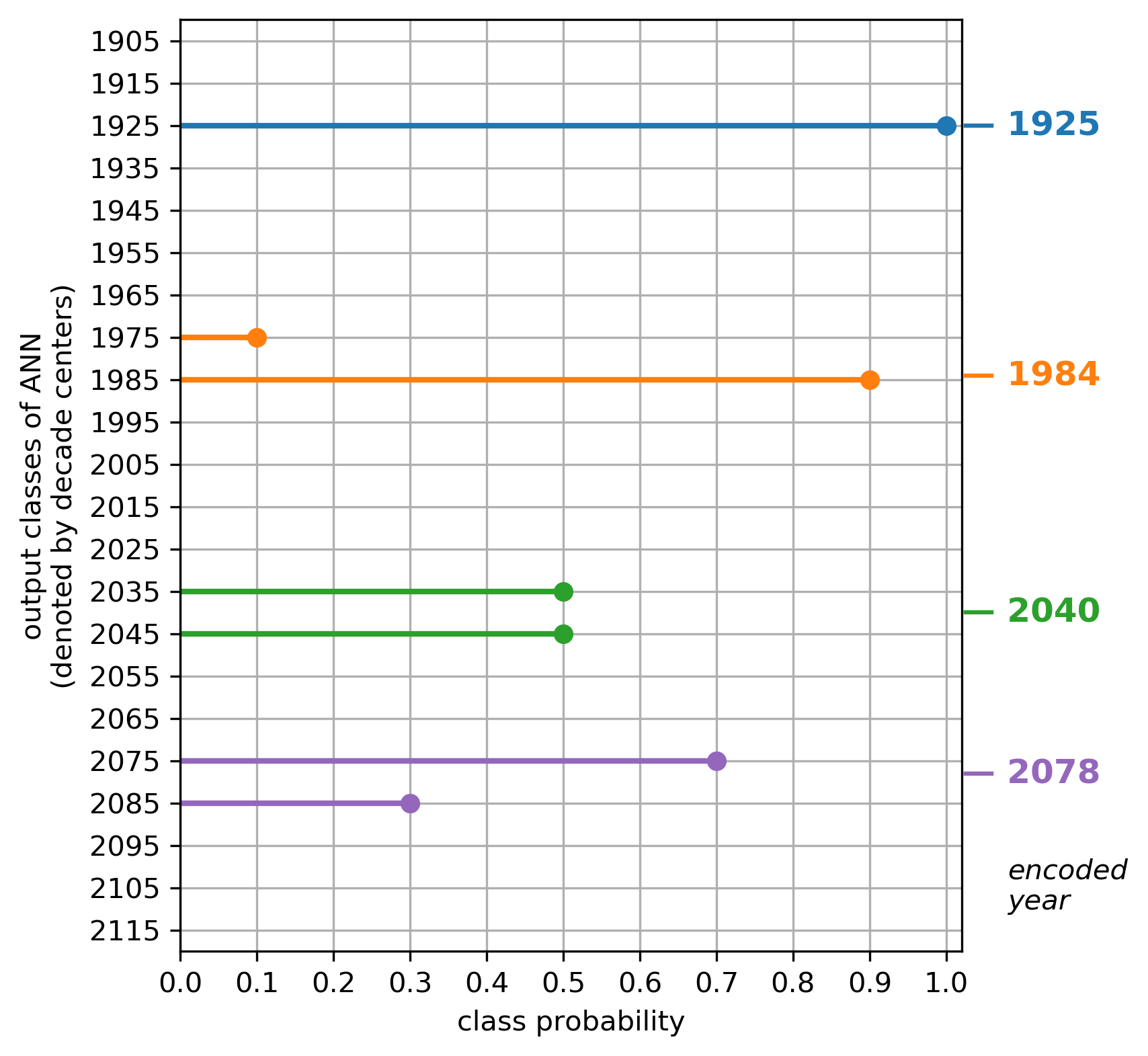}
 \end{center}
 \caption{Fuzzy classification encoding and decoding of example years. In the encoding step each colored year, 1925, 1984, 2040 and 2078, is mapped to the class probabilities indicated by the dots in the same color. For example, 1925 is encoded as probability 1.0 for class ``1925'', while 1984 is encoded as probabilities 0.1 and 0.9 for classes ``1975'' and ``1985'' respectively.  The decoding step each year can be reconstructed as the weighted sum of the decade centers, where the weights are determined by each decade's class probability. For example, 1984 results from the weighted sum $0.1 \cdot1975 + 0.9 \cdot 1985 = 1984$. }
 \end{figure}

All ANNs in this analysis have 2 hidden layers with 20 hidden units in each. This is a relatively shallow network for a typical ANN; however, our goal is to understand what the network has learned. We therefore opted for the simplest network that did not degrade accuracy. We find that increasing the number of units and/or layers does not substantially improve predictions. All units use the activation function ReLu, except for the output layer. For the output, a soft-max layer is applied before the final class probabilities are predicted.  

\subsection{Neural network training}
Each ANN is trained over the entire 1920-2099 period on 80\% of the climate model simulations and then tested on the remaining 20\%. This leads to training on 23 simulations and testing on 6 for temperature, while training on 18 simulations and testing on 4 for precipitation. Except for Figure 5, all results for a given variable utilize the same set of training/testing simulations, as well as the same neural network configuration and weight/bias initialization. This is done to make discussions more straightforward as only one ANN is analyzed at a time. The robustness of our conclusions to these choices will be discussed in Section 5.  

We trained the ANNs using the binary cross-entropy loss between the predicted class probabilities and the correct class probabilities across the training samples. Given the size of our input maps, and the small size of our output layer, the possibility of overfitting is quite large. Thus, we apply ridge regularization (L$_2$ regularization) to the weights of the first hidden layer to help reduce the chances of overfitting, and to aid in visualizing the patterns learned by the ANN (further discussed in Section 4). Ridge regularization acts to spread the importance across the inputs by adding an additional term to the cross-entropy loss that is proportional to the sum of the squared weights, which is consistent with our understanding that both temperature and precipitation exhibit substantial spatial autocorrelation. For both temperature and precipitation, the regularization parameter is 0.01. The ANN was trained using the Keras stochastic gradient descent optimizer (``SGD'') with Nesterov momentum turned on, learning rate = 0.01, momentum = 0.9, and a batch size = 32. These parameters were chosen by comparing results across a range of parameter values for each and choosing those that exhibited both high accuracies and interpretable patterns. Our results and conclusions are robust to variations in these choices. ANNs based on maps of temperature were trained for 500 iterations, while ANNs based on precipitation were only trained for 250 iterations as more iterations substantially degraded performance.

\subsection{Visualization with layerwise relevance propagation (LRP)}
A major aim of this work is to determine the patterns of forced change learned by the ANN that act as reliable indicators of the year (i.e. the class probabilities). To do this, we utilize a neural network interpretation method called ``layerwise relevance propagation'' (LRP) to determine the most relevant regions of the input maps for the ANN's prediction (e.g. Bach et al., 2015; Montavon et al., 2017). Toms et al. (2019) provide the first detailed discussion of how LRP can be used for interpretable neural networks in geoscience. We also provide the most relevant details of the method here.

LRP is a neural network interpretability method that projects the logic, or decision-making process, of a neural network back onto the original dimensions of the input. LRP traces the pathways through which information flows during the network's decision-making process for each individual sample, and shows which locations in the input image the network focuses its attention on the most (i.e. the relevance of each input pixel). LRP is implemented in the following way. Once a neural network has been trained, a sample is passed through the network to obtain a prediction (i.e. output value). This single-valued prediction is then propagated backwards to infer the relevance of each input pixel for that sample's prediction. With LRP, the output value is conserved as it is propagated backwards, which ensures that all of the information used to arrive at the network's decision is projected back onto the original input.  

Since LRP propagates only a single output value at a time, we propagate relevance only for the decade with the largest output value (i.e. probability or likelihood) predicted by the neural network, even though our fuzzy encoding requires multiple probabilities to encode a single year. Thus, the resulting relevance heatmap represents the regions of the globe that were most relevant to the neural network's confidence that the input sample belongs to that decade. Even though we propagate only the information from the decade with the highest output probability, samples from different years, e.g. 1992 and 1998 will still result in different heatmaps since the pathways through which the information flowed to generate the distributions of probabilities were different. Furthermore, we have verified that propagating all output probabilities separately (rather than just the largest) and summing their resulting relevance heatmaps leads to similar conclusions.

\subsection{Backward optimization}
Backward optimization can be used to gain a composite interpretation of the patterns contained within a trained neural network (Olah et al., 2017; Simonyan et al., 2013; Yosinski et al., 2015). Toms et al. (2019) discuss the nuances of using backward optimization for geoscience applications, and we extend its use to interpret differences between climate models and the observations. Briefly, given a trained neural network, an input sample is incrementally adjusted towards the pattern most closely associated with a user-defined prediction. This adjustment procedure utilizes a similar method that we used to update the neural network weights and biases during training (i.e. backpropagation). Rather than updating the weights and biases, however, the input is incrementally updated to minimize the difference between the user-defined desired prediction and the prediction associated with the optimized input. 

We use backward optimization to understand differences between the patterns of forced change within climate models and those within observations. As discussed within Section 3, we train neural networks to identify patterns of forced change within an ensemble of CMIP5 simulations, from which the neural network can identify the year of input maps of observed surface temperature and precipitation with reasonable accuracy. We then use backward optimization to optimize the observational maps to the networks' understanding of the climate simulations to infer biases within the climate models, the details of which are discussed within Section 6.2. During the optimization procedure, we use a learning rate of 0.001 and stop optimizing the inputs once the predicted year is correct from that point on (189 iterations for temperature, 122 iterations for precipitation). The resultant changes from optimization therefore represent the minimum change necessary to the input map in order for the neural network to correctly identify the year.

\section{Predictions based on multiple linear regression}
While this work is focused on results from a nonlinear ANN, it is informative to first discuss results using a standard linear approach. A linear approach, in particular, is useful for establishing a baseline for assessing the importance of nonlinearities when predicting with a multi-layer ANN. We begin by using all of the grid points from a simulated annual-mean surface temperature map to predict the year of the map via multi-linear regression. That is, 
\begin{equation}
\textrm{predicted year} = c + a_1x_1 + a_2x_2 + a_3x_3 + \cdots + a_{4050}x_{4050}
\end{equation}
where c denotes a constant, $x_j$ denotes the jth grid point on the globe (4050 in total) and $a_j$ denotes the regression coefficient associated with that gridpoint, or the contribution of $x_j$ to the year prediction. Furthermore, while LRP is not yet commonly used in climate science for interpreting neural networks, the general idea can be described using techniques from linear regression, providing intuition for climate scientists more familiar with this method. To make the comparison between the linear and non-linear ANN as simple as possible, we train the linear model similarly to the non-linear ANN (i.e., using backpropagation and gradient descent over 1000 iterations with a learning rate of 0.001).

Figure 3a shows the resulting predictions by this multi-linear regression model based on temperature maps, with the predicted year on the y-axis and the actual year on the x-axis. The gray dots depict the climate model simulations used for training, while the colored dots depict the simulations used for testing. This linear model appears to do an adequate job predicting the year, with most of the dots falling somewhere along the one-to-one line (which denotes a perfect prediction). To visualize these predictions, Figure 3b shows a map of the regression coefficients ($a_j$ in Eq. 1), and depicts the importance of each input grid point for the ultimate prediction of the year. This is similar to what LRP provides for nonlinear neural networks - a picture of the importance of each input unit for the final prediction. 

 \begin{figure}
 \begin{center}
 \includegraphics[width=\textwidth]{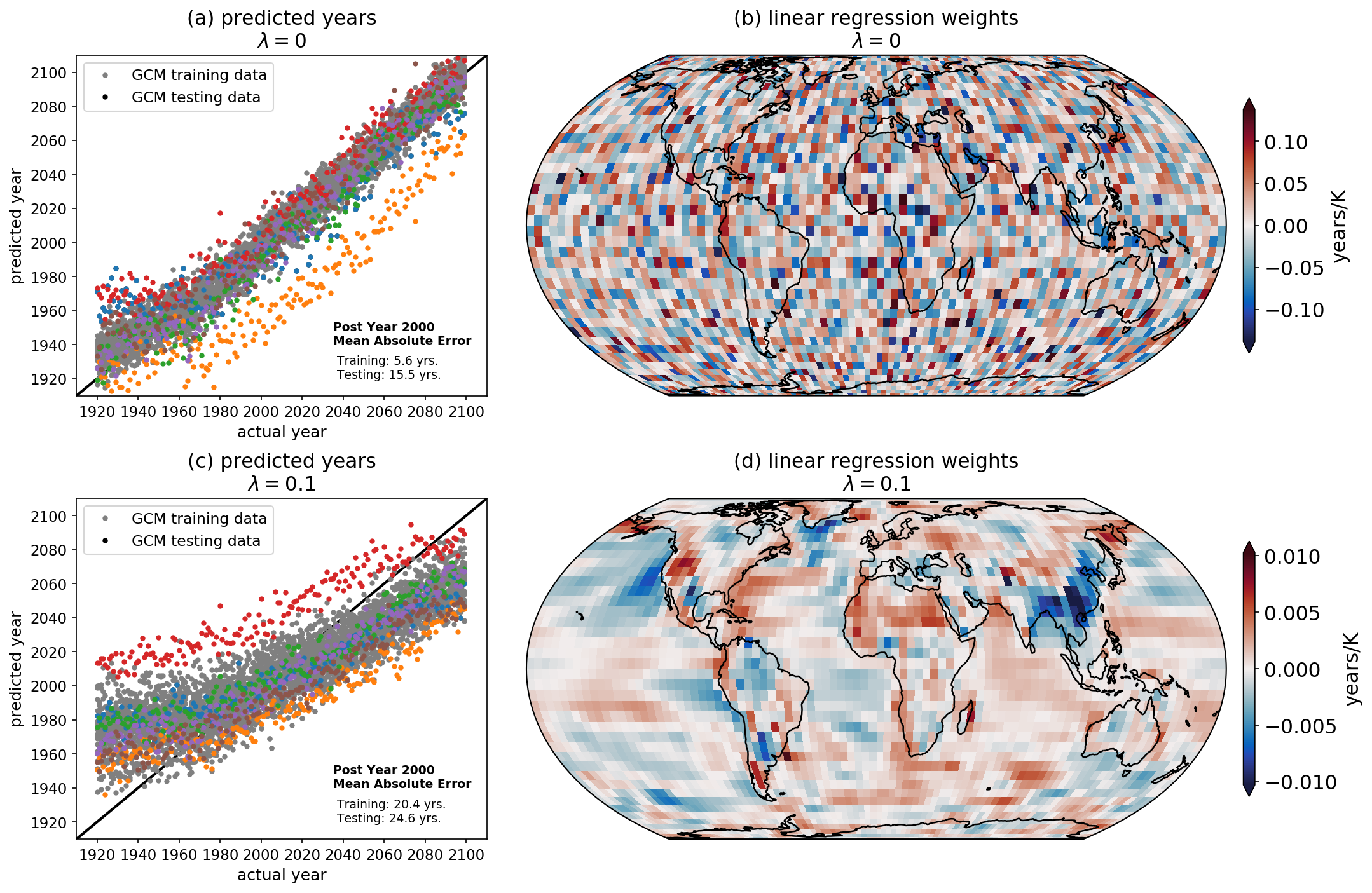}
 \end{center}
 \caption{Temperature. Predictions and regression weights from using multi-linear regression of temperature at each grid point to predict the year of the map. The upper row (a,b) uses no regularization ($\lambda=0.0$) and the lower row (c,d) utilizes L$_2$ regularization ($\lambda=0.1$).}
 \end{figure}

Although the predictions in Figure 3a generally lie along the one-to-one line, the map of regression coefficients in Figure 3b is nearly impossible to physically interpret because neighboring points often have large, opposite-signed weights. This occurs because the regression problem is under-constrained (i.e. there is a high degree of collinearity among neighboring grid points), and thus, the regression task is permitted to overfit to the noisy patterns within the temperature maps rather than the physically meaningful larger-scale patterns which are a known characteristic of atmospheric climate variability. Introducing regularization, which penalizes weights with unnecessarily large values, spreads the weights across multiple grid points, and leads to more coherent behavior between neighboring points, as seen in Fig. 3d. In other words, regularization imposes spatial auto-correlation, a known property of geophysical data, and allows us to physically interpret the learned regression weights. Warmer temperatures in western North America and northern Africa, for instance, lead the model to predict a later year, while warmer temperatures over eastern China and the eastern North Pacific drive the model to predict an earlier year. In fact, Sippel et al. (2020) apply regularized linear regression to identify a single fingerprint of external forcing in daily surface temperature maps. 

This multi-linear regression example illustrates a few key points which are useful when thinking about nonlinear ANN predictions. First, one can interpret the regression model's prediction by visualizing the importance of each input unit (i.e. each predictor grid point) for the final output. Second, L$_2$ regularization is necessary for interpreting the learned patterns, although this can come at the price of reduced accuracy in the predictions (compare Figure 3a and 3c). Since the aim of our study is to understand the patterns learned by the ANN, a small reduction in accuracy is acceptable. Furthermore, we find that L$_2$ regularization actually improves the nonlinear ANN accuracy for unseen testing data since it reduces the chances of overfitting. Third, the interpretation of the multi-linear regression prediction can be summarized in a single map that is static through time (Figure 3b,d); however, in Section 6 we show that LRP allows us to visualize the importance of a region for the ANN's prediction as a function of time.

\section{Predictions based on ANNs}
Figure 4a shows the prediction of the year by a nonlinear ANN based on input maps of surface temperature from climate model simulations. B19 showed similar panels, but here, predictions are based on the fuzzy classification scheme described in Section 3.1. As in Figure 3a,c, the gray and colored dots denote the training and testing simulations, respectively. Comparing Figure 4a with Figure 3a and 3c, it is clear that the ANN does a better job predicting the year - both for the training and testing simulations - compared to multi-linear regression. This strongly suggests that nonlinearities are important for accurate predictions. However, as discussed extensively in B19, the ANN performs poorly prior to $\sim$1960 and becomes very accurate as one moves later into the 21st Century. This is due to the increasing amplitude of forced change with time, making it easier for the ANN to identify the year amidst a background of internal variability and model disagreement over the later period.

 \begin{figure}
 \begin{center}
 \includegraphics[width=\textwidth]{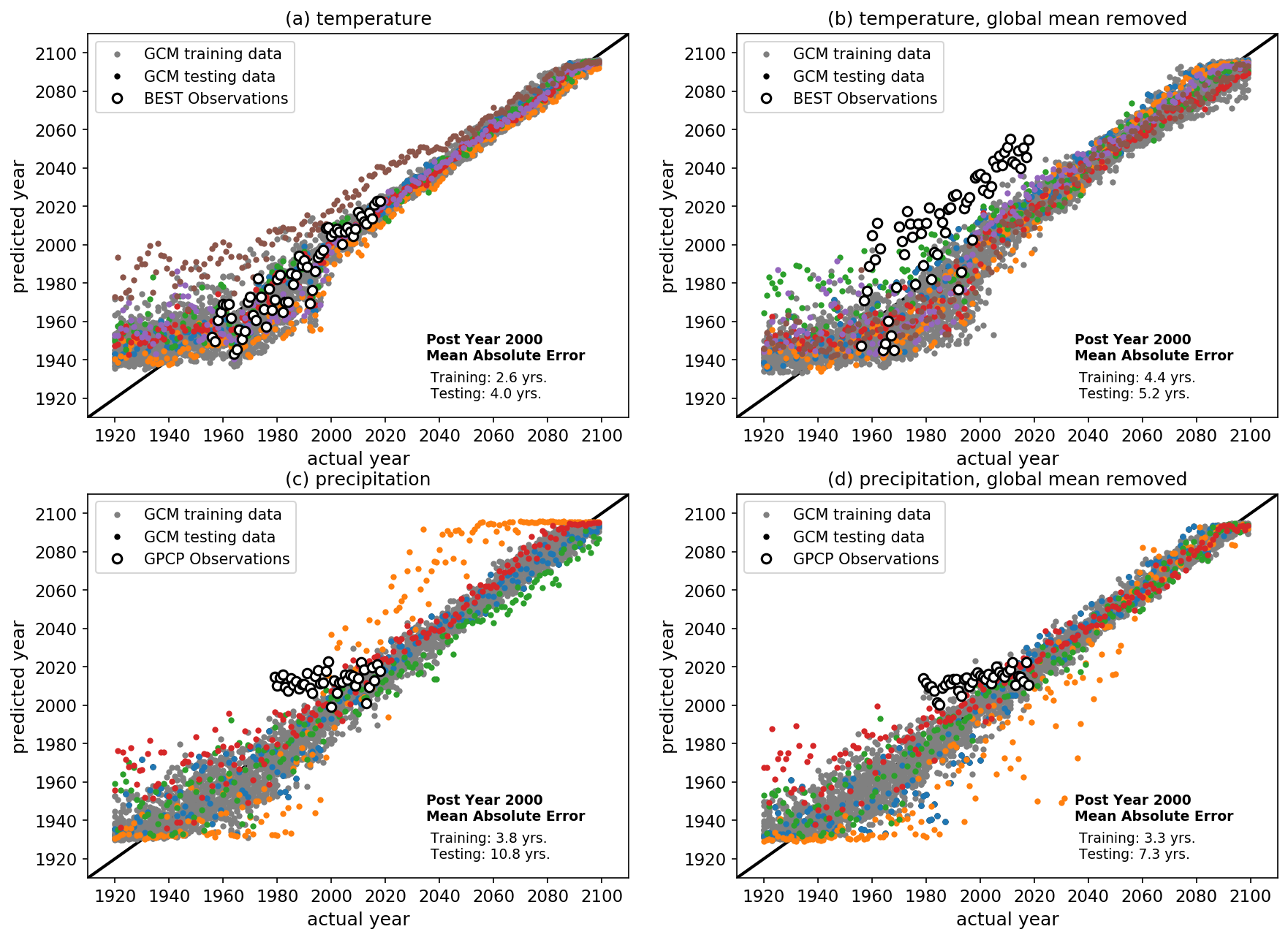}
 \end{center}
 \caption{(a) Year predicted by the neural network (y axis) versus the actual year (x axis) for (a) global maps of 2-meter temperature, (b) as in (a) but the global mean temperature has been removed from each map, (c) precipitation and (d) as in (c) but  the global mean precipitation has been removed from each map. The one-to-one line is plotted in black. Training data is shown in gray, while colors denote the different CMIP5 model simulations used for testing. The white circles denote predictions based on observed maps.}
 \end{figure}

White circles in Figure 4a depict predictions where maps of observed annual-mean surface temperature from the BEST data set are fed into the ANN trained on the climate models. Although the ANN was not trained on observed maps, it still succeeds at predicting the year when fed observed maps. This implies that the ANN is learning patterns of forced change from the climate models that are relevant for the observed climate system. As in B19, we additionally train the ANN using maps where the global mean temperature for that year has been removed. This allows us  to assess the accuracy of the ANN when it must focus on regional patterns alone. The result is shown in Figure 4b, and while the predictions spread further from the one-to-one line compared to Figure 4a, the predictions still fall  within 5 years of the true year post-2000. The biggest difference when the global mean is removed is that the predictions based on observed maps of temperature shift upward ($\sim$ 30 years later). This suggests that the regional patterns learned from the climate models may be delayed compared to what has been observed. We will explore these specific regional patterns further in Section 5.3. 

While temperature exhibits one of the most robust responses to anthropogenic emissions over the 21st Century, precipitation is primarily driven by changes in atmospheric dynamics. As a result, the  precipitation response is much less certain - with larger internal variability and less year-to-year agreement across models (Santer et al., 1994). ANN predictions of the year trained and tested on maps of annual-mean precipitation are shown in Figure 4c,d. Perhaps surprisingly, the ANN predictions for the climate model simulations  largely fall along the one-to-one line, even when the global mean has been removed.  This suggests that the ANN can identify reliable indicators of forced change in annual-mean maps of precipitation within both the 20th and 21st centuries. The predictions based on precipitation from GPCP, however, are not as successful. While the ANN largely gets the ordering of the years correct when the global mean is removed (Figure 4d), the slope of the predictions is far shallower than the one-to-one line, suggesting that the timing of reliable patterns of change differ between the observations and climate models. We revisit this discussion in Section 5.2.

While each panel of Figure 4 depicts only a single trained ANN, different ANN initializations and training/testing sets can often lead to different results. Of particular interest here is the ability of the ANN to correctly predict the year of observed maps. In Figure 5a we plot the correlation of the actual years with the predicted years based on observed maps of temperature for 21 iterations of training the ANN  (vertical orange lines). All correlations exceed 0.9, suggesting that all of the ANNs are able to discern the correct ordering of the years. When this process is repeated for input maps with the global mean removed (vertical purple lines), the correlations are reduced, as one might expect, since the ANN must rely solely on local spatial patterns of change. However, whether the global mean is retained or removed, the correlations far exceed the distribution of correlations one might expect from chance (gray histogram). An alternative metric for assessing the observational predictions is the slope of the observed year predictions, with a perfect slope being 1.0. These slopes are shown in Supp. Figure 2a, and also demonstrate that the ANN is doing much better than one would expect from chance.

 \begin{figure}
 \begin{center}
 \includegraphics[width=\textwidth]{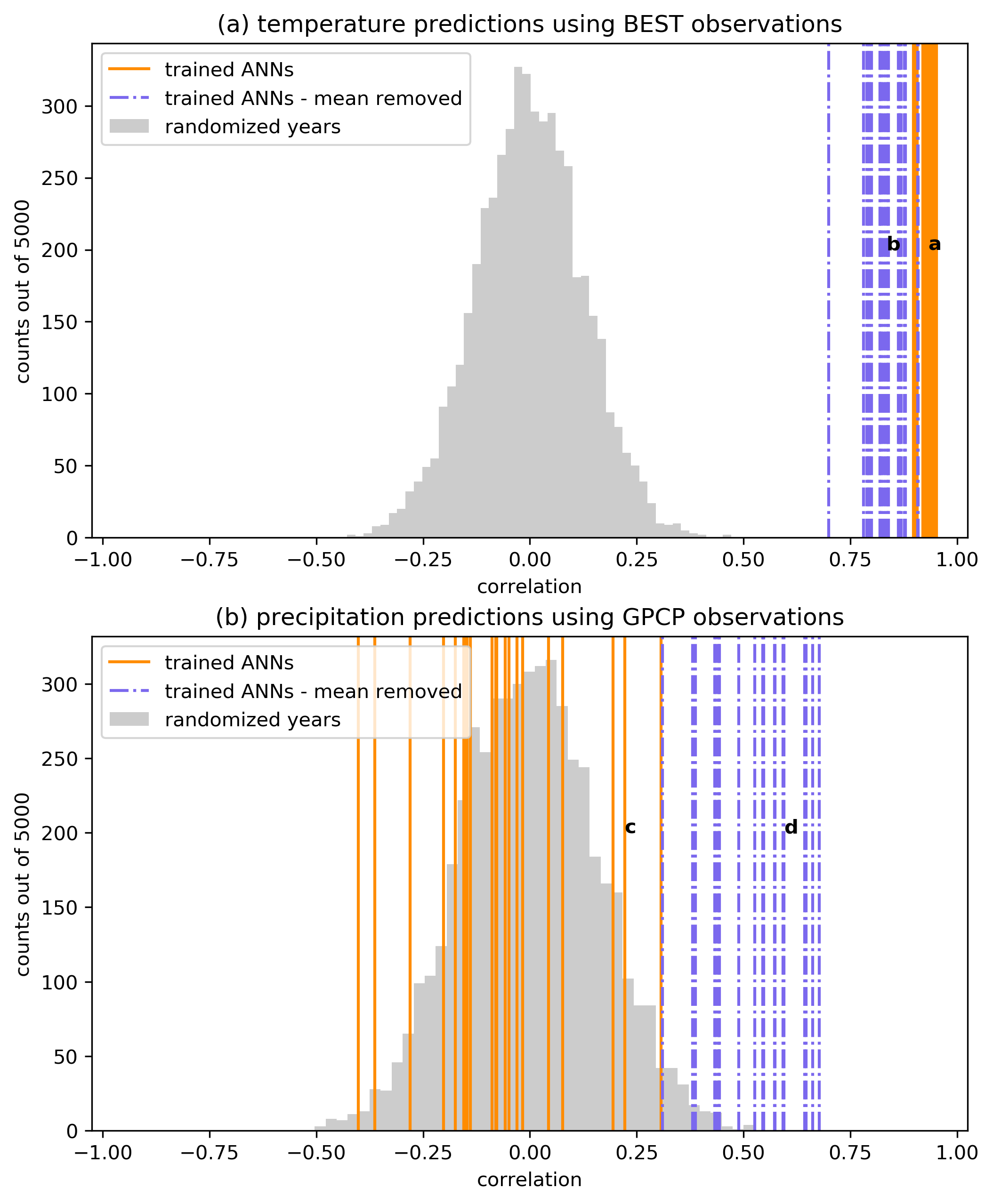}
 \end{center}
 \caption{Correlation of the actual years with the ANN-predicted years based on observed maps of (a) temperature and (b) precipitation. Different lines denote different iterations of training the ANN. (gray shading) Histogram of possible correlations between two time series with shuffled years (i.e. the range of correlations obtained when no relationship is present). Bold letters denote the iterations that are associated with the four panels of Figure 4.}
 \end{figure}

Observation-based correlations are smaller for precipitation compared to temperature (Figure 5b), consistent with the smaller signal-to-noise ratio and larger disagreement in the forced response across climate models. Unlike for temperature, the precipitation-based correlations are much larger when the global mean is removed (vertical purple lines) compared to when it is retained (vertical orange lines). In fact, most of the trained ANNs exhibit negative correlations when the mean is retained, implying a complete inability to predict the progression of years from observed maps of precipitation. The distribution of observed slopes (Supp. Figure 2b) is also better when the global mean is removed, although the slopes still fall short of 1.0. The improvement in predictions when the global mean is removed is indicative of a systematic difference between the global mean precipitation of the GPCP observations and that of each of the CMIP5 simulations (Supp. Figure 3). When the global mean is removed, the local patterns learned by the ANN trained on the climate models are more relevant for predictions of observations.

\section{Indicator patterns}
\subsection{Time varying indicators of change}
While the results in Figures 4 and 5 demonstrate the ability of an ANN to predict the year of a temperature (or precipitation) map, scientifically it is far more interesting to determine which patterns the ANN uses to identify the year. That is, which regions serve as indicators of change amidst a background of climate variability and model uncertainty? To answer this question, we apply LRP to the trained ANNs to identify the relevant regions for the ANN's predictions. As discussed previously, this is akin to making regression coefficient maps (e.g. Figure 3b,d), but instead, these relevance maps can be made for each input/prediction separately to highlight the regions of the globe that act as the most reliable indicators of the year according to the ANN.

We apply LRP to predictions from all of the training and testing simulations. Since LRP outputs a single relevance heatmap for every input/prediction, we have a total of 29 relevance heatmaps based on temperature (one per simulation) for every year from 1920-2099. Figure 6 shows the average over all heatmaps for the specified year when the predictions are deemed ``accurate'', defined as when predictions are within +/- 2 years of the true year. To increase sample sizes, we also include predictions for years within +/- 2 years of the specified year in the average as well. For example, the average relevance map for the year 2015 includes an average over all ``accurate'' predictions for maps from 2013-2017 (a total of N = 60). Since prediction accuracy largely improves as the forced signal grows in time, the number of accurate heatmaps averaged together also generally increases from the 20th to 21st Century (denoted by N in each panel).

\begin{sidewaysfigure}
 \begin{center}
 \includegraphics[width=\textwidth]{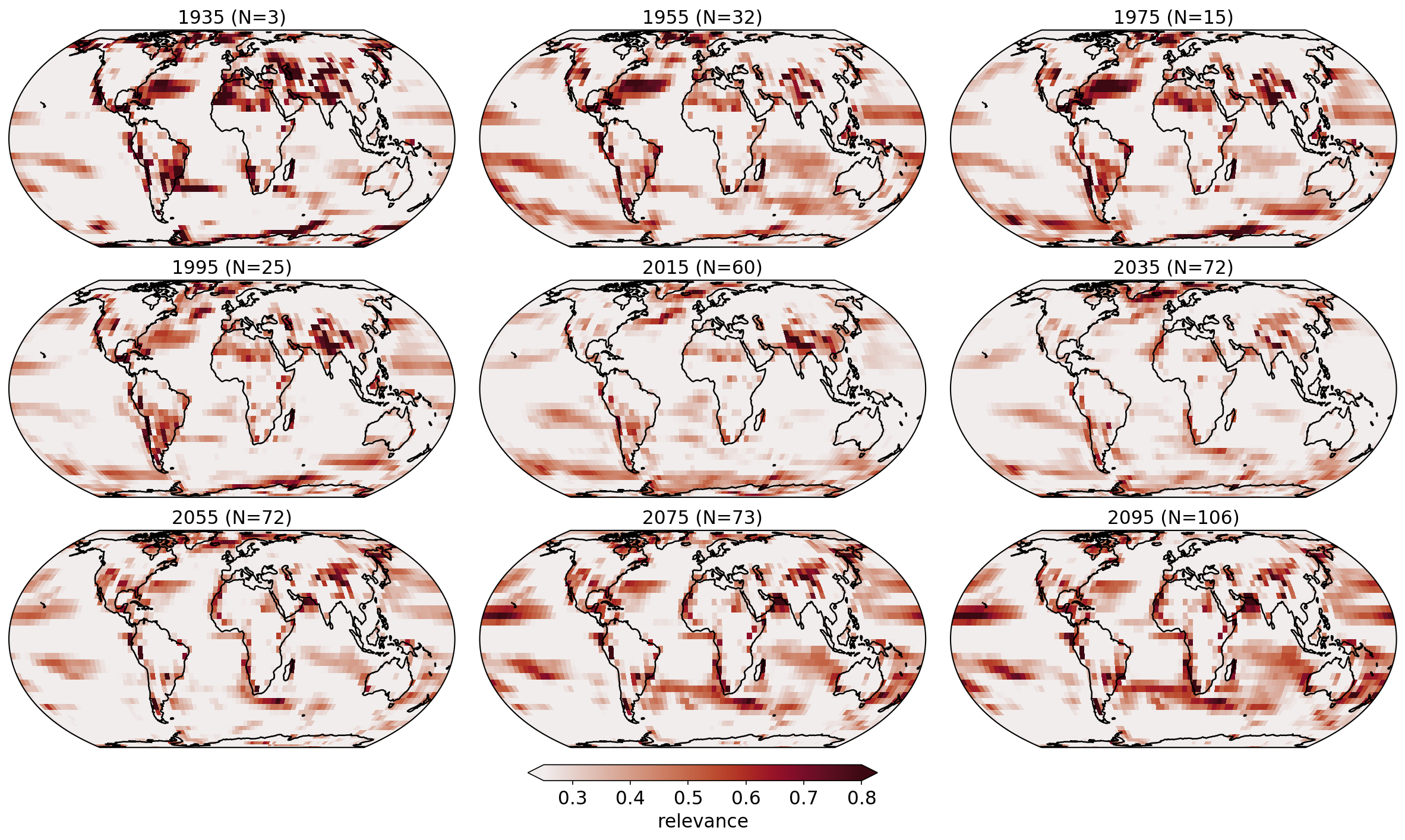}
 \end{center}
 \caption{Temperature. Layerwise relevance propagation (LRP) heatmaps for temperature input maps composited for a range of years when the prediction was deemed accurate (see text for details). The years are shown above each panel along with the number of maps used in the composites. Darker shading denotes regions that are more relevant for the ANN's accurate prediction.}
\end{sidewaysfigure}

The average LRP heatmaps in Figure 6 illustrate the most relevant regions used by the ANN (Figure 4a) to accurately predict the year of each temperature map (results for when the global mean is removed are shown in Supp. Figure 4). While akin to the regression coefficient maps in Figure 3b,d, these relevance heatmaps vary in time due to the architecture of the ANN and thus reflect the most reliable indicators of change for a particular year. The high-latitude North Atlantic exhibits large relevance over the 20th and early 21st century, while the Southern Ocean appears to increase in relevance throughout the 21st century. Eastern China lights-up as a relevant region for 1970-2020, and in fact, the multi-linear regression method (Figure 3d) also identifies eastern China as a key region for predicting the year. The difference is that the ANN allows regions to play larger roles during some decades compared to others. This is shown more clearly in Figure 7, where we plot the average relevance (as a percentile of the relevance across each input map) for eastern China and the north Arabian Sea as a function of year. While the north Arabian Sea becomes more and more relevant over time for the ANN's prediction, eastern China appears most relevant at the turn of the century. This likely reflects the strong forcing signal due to aerosols during these decades, which acts to cool the local temperatures (Fiore et al., 2015; see their Figure 4). Thus, the ANN has learned that strong cooling in eastern China relative to other regions is an indicator that the map is likely from the turn of the century. The north Arabian Sea appears to become more relevant with time because of its relatively small internal variability and so the forced signal emerges in the early 21st Century and remains strong (as shown later in Figure 9c and Supp. Figure 5c).

 \begin{figure}
 \begin{center}
 \includegraphics[width=\textwidth]{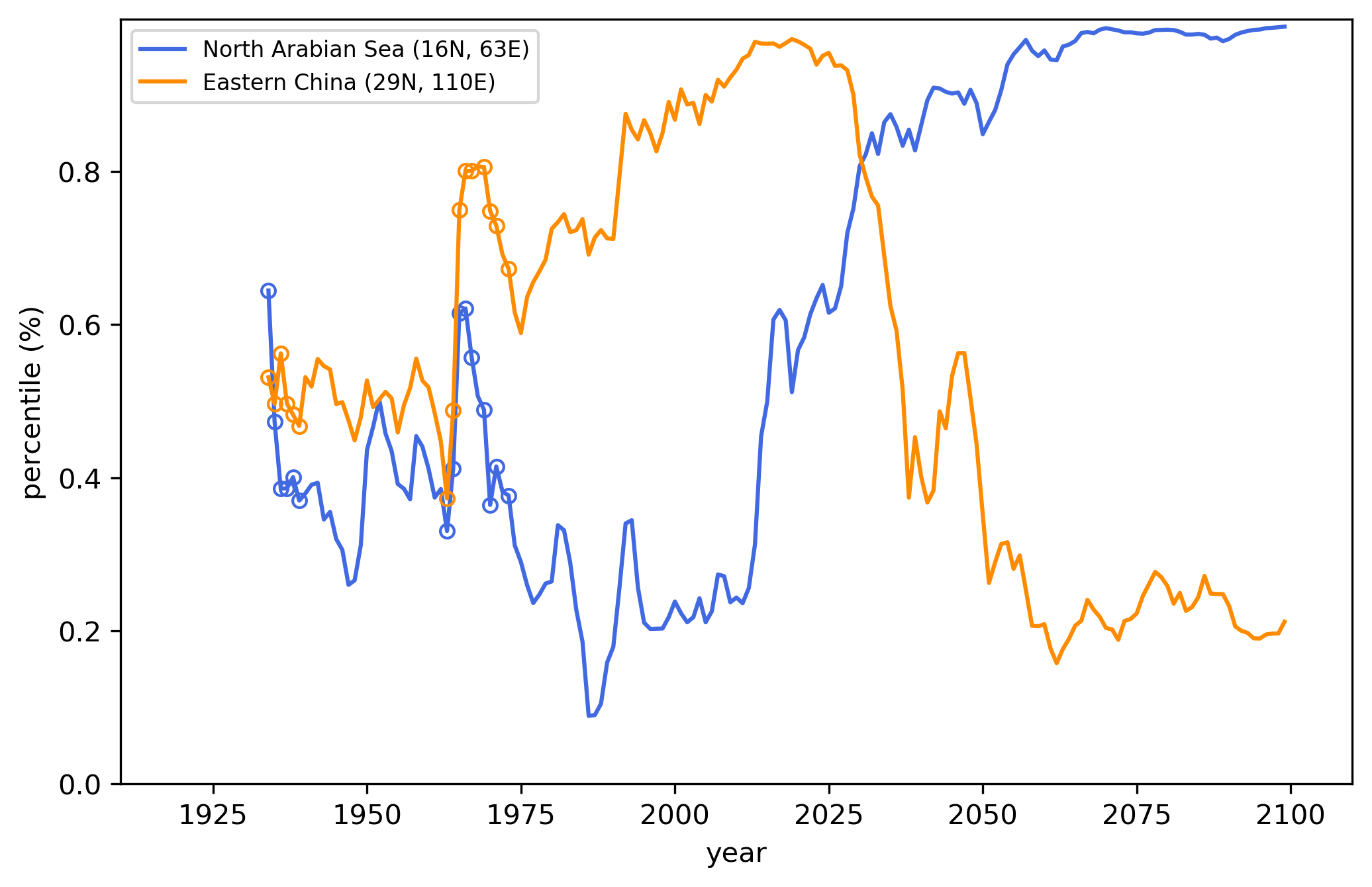}
 \end{center}
 \caption{Temperature. Average relevance percentile as a function of year for eastern China and the North Arabian sea using the temperature-based ANN. Relevance is only averaged over accurate predictions (see text for details), and averages with fewer than 10 samples are denoted with an `o'.}
 \end{figure}

Given the formulation of the LRP method, it is important to remember that the temporal evolution of a region's relevance should not be solely interpreted as its temporal forced climate response. Instead, these maps indicate the most relevant regions for a particular prediction, and so, a region may lose relevance if other regions become more relevant in later years.

Relevance heatmaps for the precipitation when the global mean is removed (i.e. Figure 4d) are shown in Figure 8. Even with L$_2$ regularization, these heatmaps appear noisier than those for temperature due to the more local nature of precipitation. Even so, relevant indicator patterns can still be seen. For example, LRP highlights Antarctica and eastern China as relevant when making accurate predictions during the 20th century. By the end of the 21st century, however, the western coasts of South America and southern Africa, as well as the Mediterranean, dominate the relevance maps. The regions highlighted by LRP signify the nonlinear, time evolution of where the signal-to-noise is large, and/or where the models agree on the response, and/or where relationships between grid points can be leveraged. 

\begin{sidewaysfigure}
 \begin{center}
 \includegraphics[width=\textwidth]{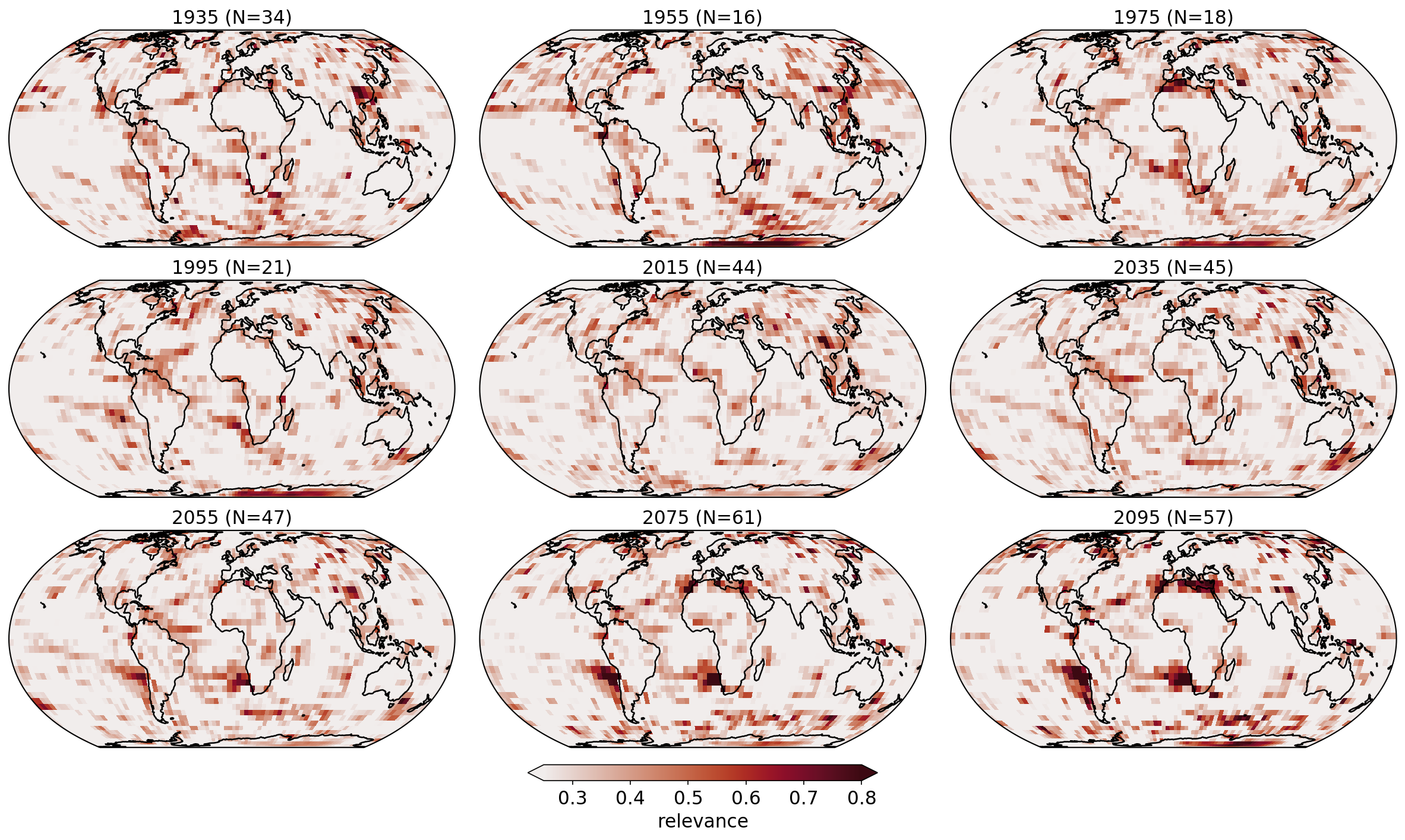}
 \end{center}
 \caption{Precipitation, global mean removed. Layerwise relevance propagation (LRP) heatmaps for precipitation input maps with the global mean removed composited for a range of years when the  prediction was deemed accurate (see text for details). The years are shown above each panel along with the number of maps composited. Darker shading denotes regions that are more relevant for the ANN's accurate prediction.}
\end{sidewaysfigure}

Given this, many of the indicator regions identified by the ANN have direct ties to more standard signal-to-noise patterns used frequently in climate science. Figure 9 shows these standard signal and signal-to-noise maps for temperature (Figure 9a,c,e) and precipitation (Figure 9b,d,f) for the turn of the century (1990-2009). Similar maps for the end of the 21st Century (2070-2099), when the forced climate change signal is much larger, are provided in Supp. Figure 5. 

Figure 9a shows the change in mean surface temperature between 1990-2009 and 1920-1949, averaged over all of the climate model simulations. This is the classic temperature change ``signal''. The well known pattern of Arctic amplification is evident, whereby the Arctic warms at an accelerated rate compared to the rest of the globe (Fyfe et al., 2013; Holland \& Bitz, 2003). Figure 9c shows the model-averaged signal-to-noise ratio, which quantifies the ratio of the signal (Figure 9a) to the year-to-year internal noise of the system. Specifically, we define this as the temperature signal for each model divided by that model's standard deviation of annual-mean temperature over the 1920-1949 period, then averaged across all models. Finally, Figure 9e provides a measure of signal-to-model disagreement, whereby the signal is defined as the model-averaged signal (Figure 9a) divided by the total spread of the signal (maximum - minimum) across the climate models. Focusing once again on the Arctic, although the signal is large (Figure 9a), the internal variability and model disagreement are too, and thus, the signal-to-noise ratios in Figure 9c,e are small. This low Arctic signal-to-noise ratio is also learned by the ANN, as seen in the LRP relevance maps in Figure 6. This is why the ANN chooses not to focus on the Arctic when making its predictions. Figure b,d,f are defined similarly but for precipitation. 

 \begin{figure}
 \begin{center}
 \includegraphics[width=\textwidth]{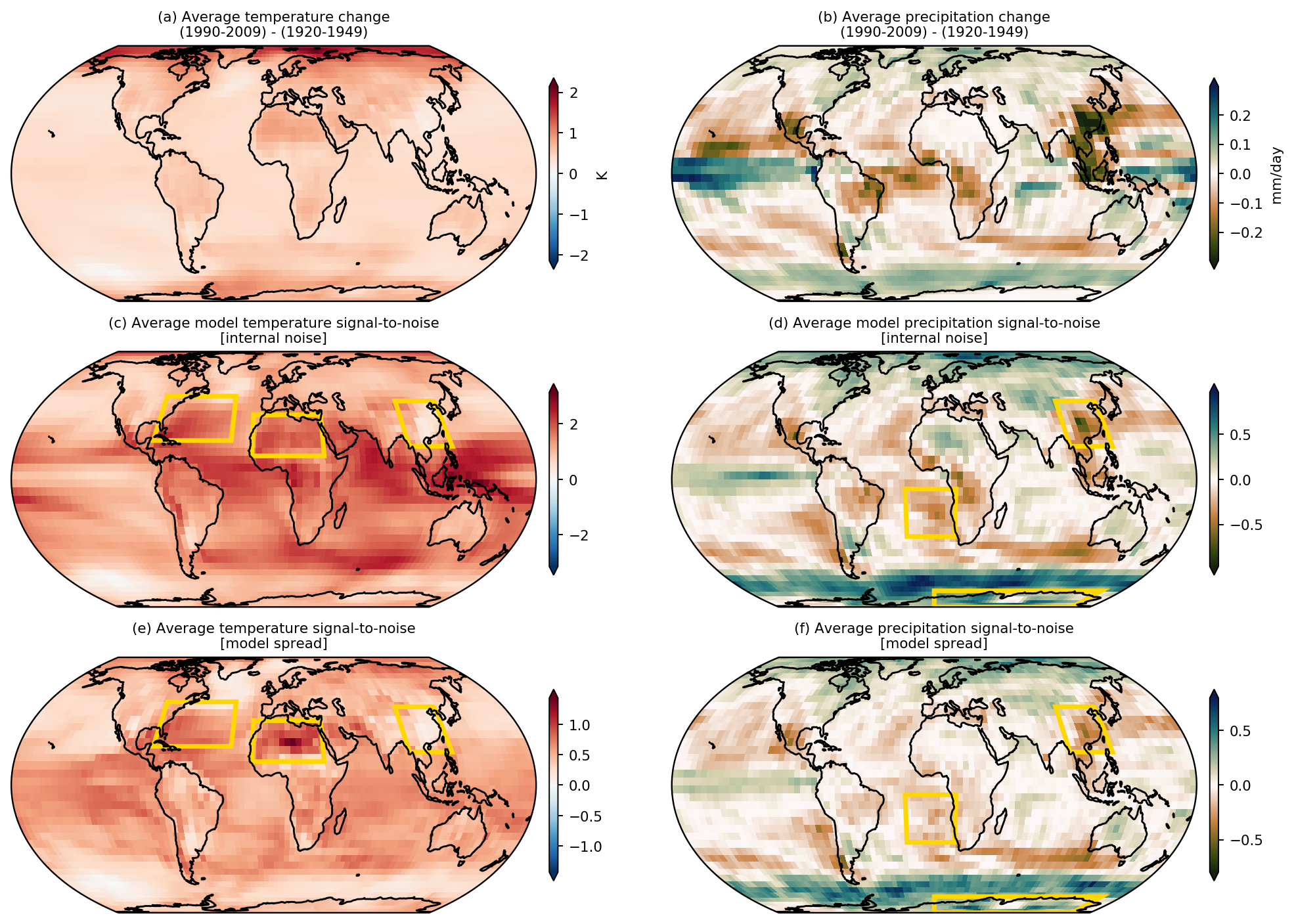}
 \end{center}
 \caption{(a) Multi-model average change in temperature between 1990-2009 and 1920-1949. (c) The average across models of each model's signal-to-noise ratio, where the signal is defined by the change in temperature and the noise is defined by the internal noise of the model (see text for details). (e) The multi-model signal-to-noise ratio, where the signal is defined by the change in temperature and the noise is defined by total spread/range of change across models. (b,d,f) as in (a,c,e) except for precipitation. Yellow boxes denote example regions which show enhanced importance using LRP in Figures 6 and 8.}
 \end{figure}

Yellow boxes in Figure 9 highlight example regions during the 1990's and 2000's that show enhanced relevance using LRP (Figures 6 and 8). For example, northern Africa is identified as important for accurate ANN predictions over the 1990s, and this region is also seen to have generally large model agreement in its response (Figure 9e). Eastern China is also identified as relevant for the ANN for both temperature and precipitation (Figures 6 and 8). For precipitation (Figure 9b,d,f), the signal, signal-to-noise and model agreement are all large there, but for temperature, the signal and signal-to-noise appears near zero. The weak temperature response (or in some cases, cooling) over eastern China, however,  compared to the warming elsewhere acts as a reliable indicator of the year. Other similar regions identified by LRP and standard signal-to-noise maps include the North Atlantic for temperature, and Antarctica/Southern Ocean for precipitation. With that said, we do not expect all of the patterns identified by LRP to appear in the signal-to-noise maps as LRP allows relationships between regional signals to be leveraged non-linearly in time, something that is not captured by a single signal-to-noise map.

\subsection{Indicator patterns in observations}
Given that LRP allows us to identify the ``reasoning'' of the ANN for each input (prediction) separately, we can use it to identify the regions that are relevant for predictions based on observations (white dots in Figure 4). Figure 10 shows the LRP relevance heatmaps when the observed temperature maps for 1997 (Figure 10b,d) and 2015 (Figure 10c,e) are fed into the ANN (see Supp. Figure 6 for examples using precipitation). Figure 10a displays the predicted probabilities for each decade output by the ANN. Although the temperature anomaly patterns are quite different between 1997 and 2015, the ANN uses similar regions for its prediction (Figure 10d,e). Namely, the largest relevance appears to be over the Southern Ocean and western coast of southern Africa, although many other regions also have non-zero relevance. Furthermore, while 1997 exhibited a large El Nino signal (warming in the eastern tropical Pacific), and 2015 had anomalously warm temperatures throughout the northern mid-to-high latitudes, neither of these regions are identified as relevant for the ANN predictions. This once again highlights that the ANN identifies the most reliable signals/regions, rather than just the largest anomalies. 

 \begin{figure}
 \begin{center}
 \includegraphics[width=\textwidth]{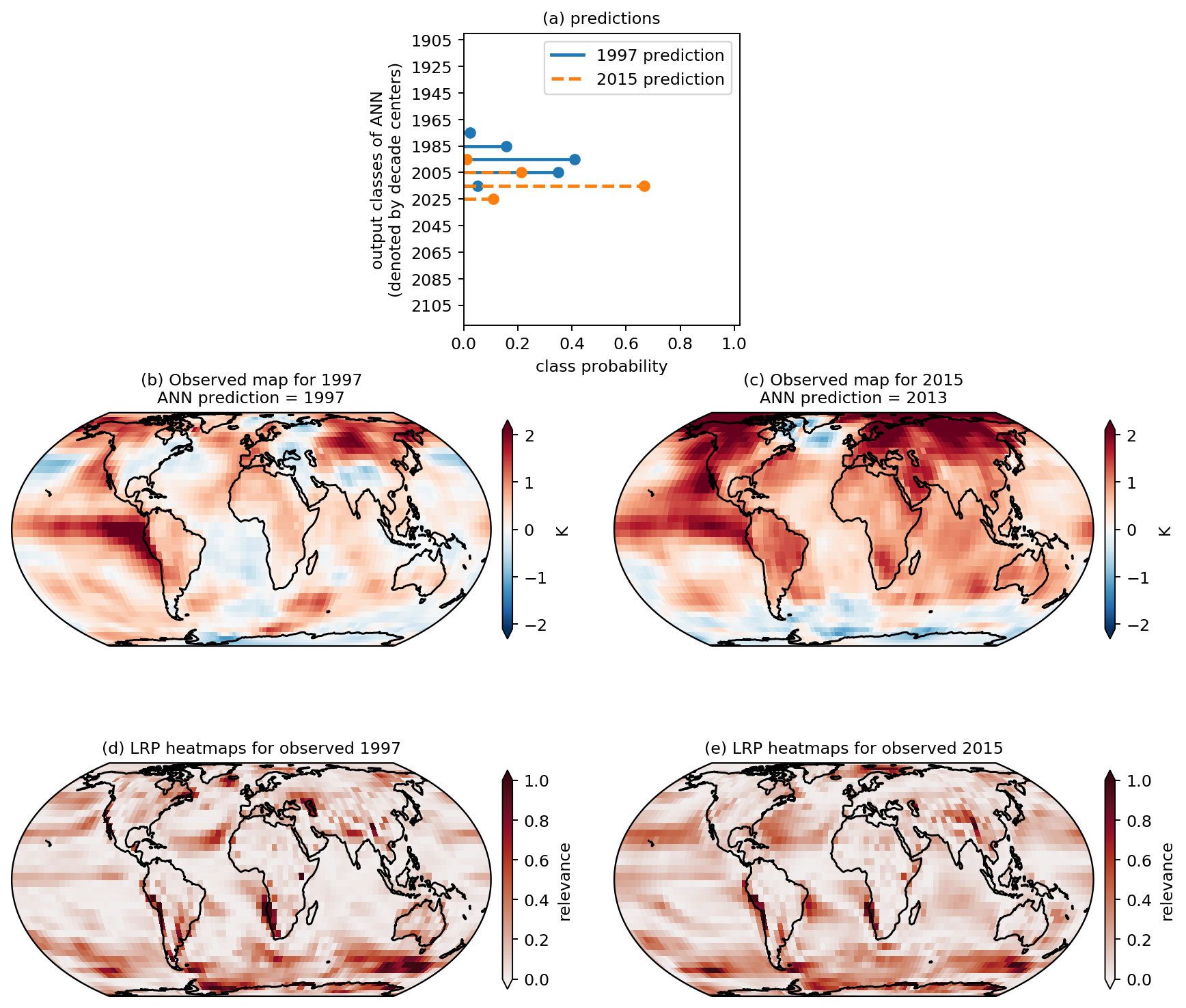}
 \end{center}
 \caption{(a) Fuzzy classification output based on observed maps of 1997 and 2015. Tick marks on the y-axis list every 2nd class for space reasons. (b,c) Observed temperature input maps plotted as anomalies from the baseline period of 1961-1990. (d,e) Layerwise relevance propagation heatmap for the ANN's year prediction.}
 \end{figure}

While the ANN predictions based on observed temperature maps are generally very good (white circles in Figure 4a), the predictions based on observed maps when the global mean is removed are shifted approximately 30 years too late (white circles in Figure 4b). Figure 11a shows the observed temperature anomalies in 1985 with the global mean removed, and the ANN incorrectly predicts the year is 2016 based on this map (31 years too far into the future). Using backward optimization (Section 3.4), we optimize the observed map (Figure 11a) to allow the ANN to make a more accurate prediction. Figure 11c shows an optimized map that leads the ANN to accurately predict 1985. While Figure 11a and 11c look very similar, their difference (Figure 11e) shows that subtle changes in the temperature patterns can improve the ANN prediction by 31 years. Figure 11g shows the same changes, but scaled by the local standard deviation of temperature (defined from linearly detrended values over the 1961-1990 baseline period). The optimized input changes reflect the changes necessary for an accurate ANN prediction, and the magnitude of these changes (either in physical units or standard deviations) correspond to  the threshold at which the optimized signal becomes identifiable above the noise.

 \begin{figure}
 \begin{center}
 \includegraphics[width=\textwidth]{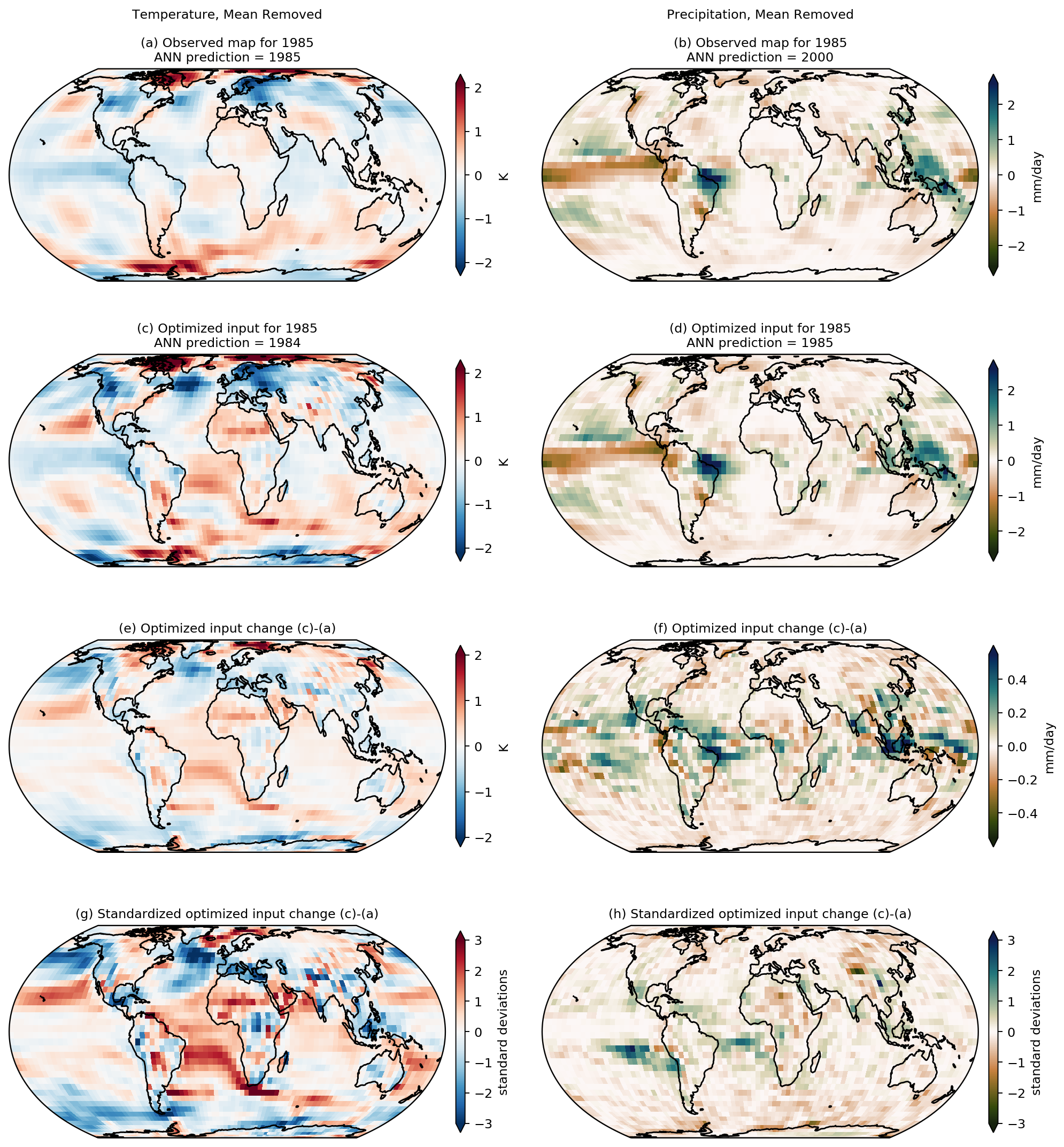}
 \end{center}
 \caption{Temperature and precipitation, global mean removed. (a) Observed temperature anomaly maps with the global mean removed, plotted as anomalies from the baseline period of 1961-1990. (c) Optimized input map determined using backward optimization. (e) Difference between (c) and (a). (g) As in (e) but standardized by the local standard deviation, defined from the detrended values over the baseline period. (b,d,f,h) Similar panels but for observed precipitation anomaly maps with the global mean removed, plotted as anomalies from the baseline period of 1979-1999.}
 \end{figure}

In a general sense, Figure 11g shows that cooling the continents and North Pacific ocean and warming the rest of the oceans in 1985 would lead the ANN to a much more accurate prediction. The concept of cooling or warming the observed globe seems rather odd since the observed map is what actually occurred. However, the ANN was trained on climate model simulations, and so, from the point-of-view of the ANN, it is the observations that need to be adjusted. If we change our framing, we can instead view Figure 11g as highlighting the fact that the climate models upon which the ANN was trained are too cold over land compared to the oceans. That is, this method has extracted a critical model bias in regional patterns of warming in the 1980s. To support the robustness of this result, Supp. Figure 7 shows that optimizing the observed 2015 map (rather than 1985) extracts a similar climate model bias - namely that the land does not warm fast enough relative to the oceans in climate model simulations (Supp. Figure 7f,h).

The right column of Figure 11 shows a similar analysis but for the observed 1985 precipitation map, where the global mean has been removed (Figure 11b). As for temperature, the ANN predicts too late of a year for this input map, predicting the year 2000 for the 1985 observed map. Backward optimization leads to the optimized map shown in Figure 11d, and when fed this optimized map, the ANN is able to predict  the correct year of 1985. Figure 11h shows the optimized changes (in local standard deviation), and we see that the optimized map has increased precipitation anomalies off of the coast of South America, southern Africa, and eastern Antarctica, and decreased precipitation anomalies over northern Africa and central Asia. Once again, these changes can be interpreted as regions where climate model simulations are too wet (blue/green shading) or too dry (brown/orange shading) relative to the GPCP observations. Supp. Figure 8f,h shows that the same regional biases are extracted when one optimizes the observed 2005 precipitation map, suggesting these biases are present for multiple decades.

\section{Conclusions}
We identify reliable indicator patterns of forced change within annual-mean surface temperature and precipitation maps from climate model simulations using artificial neural networks (ANNs) and two powerful visualization methods, layerwise relevance propagation and backward optimization. The indicator patterns vary through time, and the ANN captures the nonlinear, time evolution of the signal-to-noise ratio and model agreement by leveraging relationships between grid points. Since layerwise relevance propagation identifies the regions that are most relevant for a given prediction, we apply it to input maps of observational data that were not used during training of the ANN. We find, for example, that the ANN identifies the Southern Ocean as a reliable indicator of forced change within the observational record. Finally, we use backward optimization to identify the relevant regions where climate models are most different from observations for any given year. For example, temperature results show that models are too cold over the land and too warm over the oceans, while results for precipitation suggest that models are too wet off the western coasts of South America and Africa.

While previous work by Barnes et al. (2019) demonstrated that ANNs are capable of identifying patterns of forced change in climate model simulations, they did not present the patterns themselves due to the complexity of visualizing the nonlinear decision making process of an ANN. Since then, neural network visualization tools developed by the computer science community have been introduced to the geosciences (e.g. McGovern et al., 2019; Toms et al., 2019), and allow for visualization and interpretation of the fully nonlinear ANN. Thus, while this work highlights their use for visualizing forced patterns of change, we suggest that it is likely the first of many to demonstrate the profound ability of neural networks and their visualization methods to extract climate patterns from the noise.

\acknowledgments

Work was supported by NSF CAREER AGS-1749261 (EAB), NSF AGS-1445978 (IE), both under the Climate and Large-scale Dynamics program, and DOE grant DE-FG02-97ER25308 (BT). Data and code access: http://www.cesm.ucar.edu/projects/community-projects/LENS/data-sets.html (CESM); https://esgf-node.llnl.gov/projects/cmip5/ (CMIP5); http://berkeleyearth.org/data/ (BEST); https://www.esrl.noaa.gov/psd/data/gridded/data.gpcp.html (GPCP). We acknowledge the World Climate Research Programme's Working Group on Coupled Modelling, which is responsible for CMIP, and we thank the climate modeling groups for producing and making available their model output. For CMIP the U.S. Department of Energy's Program for Climate Model Diagnosis and Intercomparison provides coordinating support and led development of software infrastructure in partnership with the Global Organization for Earth System Science Portals. All data used in this study is publicly available and referenced throughout the paper.


%
%
\nocite{*}
\bibliography{References_from_lrpForcedPatterns_MAIN.bib}

%
%
%
%
%

\end{document}